\documentclass[a4paper,11pt]{article}
\usepackage{aaskaiid}
\usepackage{orcidlink}
\usepackage{xcolor}
\usepackage{tikz}
\usepackage{cancel}
\usepackage{bm}

\usetikzlibrary{shapes.geometric, backgrounds, arrows, calc, fit}
\tikzstyle{process} = [rectangle, rounded corners, minimum width=3cm, minimum height=1cm, text centered, draw=black]
\tikzstyle{arrow} = [thick,->,>=stealth]

\title{Using SKAO to Understand the Clustering of Gravitational Wave Sources}
\ShortTitle{SKAO cross correlation with GW sources}

\author[1,2]{Michele Bosi \orcidlink{0009-0000-8215-6698}}
\ShortName{Bosi et al.}
\emailAdd{mbosi@sissa.it}
\author[3]{Sarah Libanore \orcidlink{0000-0002-2284-9190}}
\emailAdd{libanore@bgu.ac.il}
\author[4,5,6]{Nicola Bellomo \orcidlink{0000-0002-4375-705X}}
\emailAdd{nicola.bellomo@unipd.it}
\author[4]{Caterina Scarpel}
\emailAdd{caterina.scarpel@studenti.unipd.it}
\author[4,5]{Federico Semenzato \orcidlink{0009-0006-0219-6192}}
\emailAdd{federico.semenzato.1@phd.unipd.it}
\author[4,5,6]{Alvise Raccanelli \orcidlink{0000-0001-6726-0438}}
\emailAdd{alvise.raccanelli.1@unipd.it}
\author[4,5,6]{Michele Liguori \orcidlink{0000-0002-0383-2254}}
\emailAdd{michele.liguori@unipd.it}

\affiliation[1]{Scuola Internazionale Superiore di Studi Avanzati, Via Bonomea 265, 34136 Trieste, Italy}
\affiliation[2]{Department of Physics, University of Trento, Via Sommarive 14, 38123 Povo (TN), Italy}
\affiliation[3]{Department of Physics, Ben-Gurion University of the Negev, Be'er Sheva 84105, Israel}
\affiliation[4]{Dipartimento di Fisica e Astronomia ``G. Galilei'', Universit\`a degli Studi di Padova, via Marzolo 8, I-35131 Padova, Italy}
\affiliation[5]{INFN, Sezione di Padova, Via Marzolo 8, I-35131, Padova, Italy}
\affiliation[6]{INAF - Osservatorio Astronomico di Padova, Vicolo dell'Osservatorio 5, I-35122 Padova, Italy}

\abstract{
Coalescing Binary Black Holes (BBHs) trace the Large-Scale Structure (LSS) of the Universe, and their clustering properties can be extracted from Gravitational Wave (GW) data.
Next-generation detectors, such as the Einstein Telescope and Cosmic Explorer, will enable statistical studies of GW sources thanks to the massive number of detected events.
However, such events will still suffer from significant instrumental and theoretical uncertainties.
Cross-correlating GW maps with other LSS surveys provides a promising strategy to mitigate these limitations.
The SKA-Mid intensity mapping and radio continuum surveys offer ideal datasets for cross-correlation studies with GWs (SKAO$\times$ET2CE).
Their wide sky coverage and deep redshift sensitivity will allow precise probing of the epochs and environments where stellar BBHs form most efficiently.
In this chapter, we forecast the potential of cross-correlation angular power spectra to extract information on the distribution and clustering properties of GW events.
First, we model the number density and bias of three independent tracers: GW sources, neutral hydrogen intensity maps, and radio galaxies.
We estimate the constraining power of SKA-Mid$\times$ET2CE on the GW clustering bias, which carries information on the origin of GW progenitors, e.g., whether they formed through stellar evolution or are primordial black holes.
Finally, we develop a semi-analytic model for GW events hosted by SKAO galaxies as a function of the time-delay distribution between the binary formation and merger, which is still largely uncertain to date.
We forecast the signal-to-noise ratio of their cross-correlation with SKA-Mid, and demonstrate that SKA-Mid$\times$ET2CE will foster our understanding of the time-delay distribution.
}

\begin{document}
\newcommand{\actaa}{Acta Astron.} % Acta Astronomica
\newcommand{\araa}{ARA\&A} % Annual Review of Astron and Astrophys
\newcommand{\aar}{A\&ARv} % Astrononmy \& Astrophysics Review
\newcommand{\aapr}{A\&ARv} % Astronomy\&Astrophysics Reviews
\newcommand{\ab}{Astrobiol.} % Astrobiology
\newcommand{\aj}{AJ} % Astronomical Journal
\newcommand{\apj}{ApJ} % Astrophysical Journal
\newcommand{\apjl}{ApJL} % Astrophysical Journal, Letters
\newcommand{\apjs}{ApJSS} % Astrophysical Journal, Supplement
\newcommand{\ao}{Appl. Opt.} % Applied Optics
\newcommand{\apss}{Astro. \& Space Sci.} % Astrophysics and Space Science
\newcommand{\aap}{A\&A} % Astronomy and Astrophysics
\newcommand{\aaps}{A\&AS.} % Astronomy and Astrophysics, Supplement
\newcommand{\baas}{Bull. Am. Astron. Soc.} % Bulletin of the AAS
\newcommand{\caa}{Chinese A\&A} % Chinese Astronomy and Astrophysics
\newcommand{\cjaa}{Chinese J. A\&A} % Chinese Journal of Astronomy and Astrophysics
\newcommand{\cqg}{Class. Quantum Gravity} % Classical and Quantum Gravity
\newcommand{\gal}{Galaxies} % Galaxies
\newcommand{\gca}{Geo. Cosmo. Acta} % Geochimica Cosmochimica Acta
\newcommand{\icarus}{Icarus} % Icarus
\newcommand{\jcap}{JCAP} % Journal of Cosmology and Astroparticle Physics
\newcommand{\jgr}{J. Geophys. Res.} % Journal of Geophysics Research
\newcommand{\jgrp}{J. Geophys. Res. Planets} % Journal of Geophysics Research: Planets
\newcommand{\jqsrt}{J. Quant. Spectrosc. Radiat. Transf.} % Journal of Quantitiative Spectroscopy and Radiative Transfer
\newcommand{\memsai}{Mem. SAIt} % Mem. Societa Astronomica Italiana
\newcommand{\mnras}{MNRAS} % Monthly Notices of the RAS
\newcommand{\nat}{Nature} % Nature
\newcommand{\nastro}{Nat. Astron.} % Nature Astronomy
\newcommand{\ncomms}{Nat. Commun.} % Nature Communications
\newcommand{\nphys}{Nat. Phys.} % Nature Physics
\newcommand{\na}{New Astron.} % New Astronomy
\newcommand{\nar}{New Astron. Rev.} % New Astronomy Review
\newcommand{\physrep}{Phys. Rep.} % Physics Reports
\newcommand{\pra}{Phys. Rev. A} % Physical Review A: General Physics
\newcommand{\prb}{Phys. Rev. B} % Physical Review B: Solid State
\newcommand{\prc}{Phys. Rev. C} % Physical Review C
\newcommand{\prd}{Phys. Rev. D} % Physical Review D
\newcommand{\pre}{Phys. Rev. E} % Physical Review E
\newcommand{\prx}{Phys. Rev. X} % Physical Review X
\newcommand{\prl}{Phys. Rev. Let.} % Physical Review Letters
\newcommand{\psj}{Planet. Sci. J.} % Planetary Science Journal
\newcommand{\planss}{Planet. Space Sci.} % Planetary Space Science
\newcommand{\pnas}{Proc. Natl Acad. Sci. USA} % Proceedings of the US National Academy of Sciences
\newcommand{\procspie}{Proc. SPIE} % Proceedings of the SPIE
\newcommand{\pasa}{PASA} % Publications of the Astron.  Soc. of Australia
\newcommand{\pasj}{PASJ} % Publications of the Astron.  Soc. of Japan 
\newcommand{\pasp}{PASP} % Publications of the Astron.  Soc. of the Pacific
\newcommand{\rmxaa}{RMXAA} % Revista Mexicana de Astronomia y Astrofisica
\newcommand{\sci}{Science} % Science
\newcommand{\sciadv}{Sci. Adv.} % Science Advances
\newcommand{\solphys}{Sol. Phys.} % Solar Physics
\newcommand{\sovast}{Soviet Ast.} % Soviet Astronomy
\newcommand{\ssr}{Space Sci. Rev.} % Space Science Reviews
\newcommand{\uni}{Universe} % Universe

\maketitle

%%%%%%%%%%%%%%%%%%%%%%%%%%%%%%%%%%%%%%%%%%%%%%%%%%%%%%%%%%%%%%%%%%%%%%%%%%%%%%%%%%%%%%%%%%%%%%%%%%%%%%%%%%%%%%%%%%

\section{Introduction}

Cross-correlations between multiple tracers are widely employed in Cosmology as a way to reduce the noisiness of measurements, thus enhancing the statistical significance of the detected signal, and allowing for the detection of minor physical effects that might be hidden in a single-tracer analysis.
To our knowledge, \cite{Seljak:2008xr} is the first that discusses the critical advantages of performing a multi-tracer analysis to mitigate the impact of cosmic variance, eliminate shot-noise contributions, and showcase the improvement on the cosmological parameters confidence regions.
The core idea of this work, and many others that followed, see, e.g.,~\citep{McDonald_2009, Hamaus:2011dq, Abramo_2015}, is that each tracer provides an independent sample of the underlying Dark Matter field.
Thus, by combining the clustering information of multiple tracers, we effectively probe multiple times poorly sampled modes with a characteristic size of our galaxy survey.

Since different tracers are inherently characterized by different biasing properties, multi-tracer analyses prove to be extremely effective in disentangling the effect of cosmological parameters, acting in the same fashion on all tracers, from individual tracers properties due to their unique formation history.
Regarding this last point, cross-correlation techniques already demonstrated their ability to characterize the distribution, environment, and nature of poorly constrained tracers, such as in the context of the clustering-based redshift analysis pioneered by~\cite{Newman:2008mb, McQuinn:2013ib, Menard_2014}.
In these early works, cross-correlations between spectroscopic and photometric galaxy surveys allowed for the assignment of spectroscopic redshift to photometric sources, as in the case of the Dark Energy Survey in~\cite{DES:2017rfw,Cawthon_2022}, and the Sloan Digital Sky Survey (SDSS) in~\cite{Rahman:2014lfa}.
Other fields have also benefitted from the use of the clustering-based redshift, showing that cross-correlations can be used to reconstruct the tomographic information of observables that are integrated over broad observational bands.
For instance, \cite{Cheng:2021wex, Schmidt:2014jja, Chiang:2018miw, Libanore:2024wvv} constrained the redshift evolution of the extra-galactic background light using its cross-correlation with spectroscopic galaxy surveys, while~\cite{Kovetz:2016hgp} demonstrated how a clustering-based redshift analysis between photometric and radio surveys has the potential of constraining exotic cosmological scenarios.

The performance of a cross-correlation analysis is heavily dependent on the characteristics of the most precise of the two Large-Scale Structure (LSS) surveys.
In this sense, the Square Kilometer Array Observatory (SKAO,~\cite{Carilli:2004nx, Braun:2015zta}), with its unprecedented sensitivity, sky coverage, and redshift depth, offers a unique opportunity.
In this chapter we focus on the SKA-Mid telescopes, which cover the redshift range~$z\in[0,3]$ via their radio frequency bands ($\nu_{\rm obs}\in[0.35,15.4]\,{\rm GHz}$), span almost half of the sky, and represent an ideal partner to study the Universe around Cosmic Noon, when star formation reaches its maximum.\footnote{
During the AA* stage, SKA-Mid telescopes will observe only in four frequency bands, i.e.,~$1,2,5\mathrm{a},5\mathrm{b}$.
Although still planned, Band 3 ($\nu_{\rm obs}=[1650,3050]\,{\rm MHz}$) and Band 4 ($\nu_{\rm obs}=[2800,5180]\,{\rm MHz}$) are expected to be deployed once funding becomes available.
Since they remain part of the AA4 Design Baseline, we include them in our work.}

On the other hand, cross-correlation analyses yield the greatest benefit when the two LSS tracers originate from distinct and independent formation channels. This insight has motivated the idea of combining electromagnetic measurements with Gravitational-Wave (GW) surveys.
Over the course of the past decade and its four observational runs, the LIGO-Virgo-KAGRA Collaboration has detected hundreds of GW events~\citep{LIGOScientific:2016dsl, LIGOScientific:2018mvr, LIGOScientific:2020ibl, LIGOScientific:2021usb, KAGRA:2021duu, LIGOScientific:2025slb} produced by the merger of Binary Black Hole (BBH) systems.
These compact objects are expected to be created at the end of the stellar cycle of very massive stars; thus, we expect them to trace the regions of our Universe that underwent an intense star-forming epoch.
Meanwhile, star-forming galaxies (SFGs) are observable in the radio band through their free-free and synchrotron emission. As a result, the SKAO has the ability to detect a large population of SFGs, offering a unique opportunity, via cross-correlation, to test whether these galaxies are indeed the place where BBH systems are born and have merged.
Moreover, the SKAO is expected to be fully operational by the time next-generation GW interferometers, such as the Einstein Telescope (ET,~\cite{Punturo:2010zz, ET:2025xjr}) and the Cosmic Explorer (CE,~\cite{Evans:2021gyd}), begin observations, making this synergy even more compelling and opening the possibility of cross-correlating both new and archival SKAO data with GW observations.

The seminal work of~\cite{Raccanelli_2017} showcased the natural synergy between GW and radio surveys datasets to constrain the relationship between BBHs and their hosts.
Later, \cite{Scelfo:2021fqe} demonstrated how to reconstruct the redshift distribution of GWs from BBH mergers by employing the clustering-based redshift technique in cross-correlation with high-frequency resolution HI intensity maps.
Additionally, \cite{Zazzera:2025ord} has recently shown how GW and radio cross-correlations can be exploited to test General Relativity on scales comparable with that of the cosmological horizon.
Therefore, in this chapter, we aim to characterize the potential gain of information that a cross-correlation analysis with SKAO Telescopes might bring to the GW community in terms of understanding of GW sources and their clustering properties.

To this end, we focus on two anticipated SKA-Mid surveys: one tracing radio-continuum galaxies (RC) and the other mapping neutral hydrogen intensity (HI-IM). While the RC survey will be conducted using an interferometric setup, the HI-IM survey will employ a single-dish configuration, following the pioneering approach of MeerKLASS~\citep{Santos_2017:meerkl_im}.
To model the redshift-dependent number density and bias of RC sources, we exploit \texttt{T-RECS} catalogs. For HI-IM instead, we rely on the semi-analytical model in~\cite{Crighton_2015, Battye_2013}.
In the first part of our analysis, we demonstrate that SKAO’s contribution will be essential for obtaining informative constraints on the GW clustering bias, enabling us to break degeneracies between cosmological and astrophysical parameters. Constraining the GW bias will also improve our understanding on the nature of GW progenitors, including the possibility of identifying signatures of more exotic scenarios, such as the presence of primordial black hole binaries (PBHs).

In the second part of this chapter, we take a further step and introduce a novel approach to constraining the BBH time-delay distribution, namely the probability distribution of the time interval between the formation and merger of a BBH system. This quantity depends on the binary formation channel but remains highly uncertain, even from a theoretical point of view (see, e.g.,~\cite{Fishbach:2021mhp}). With our analysis, we show for the first time that cross-correlating GWs with SKAO RC galaxy observations can yield strong constraints on the time-delay distribution.

The chapter is structured as follows.
In Sec.~\ref{sec:LSS_with_SKAO}, we summarize the formalism of the cross-angular power spectrum in the context of SKAO science and describe the redshift distribution and bias of the SKAO tracers adopted in our analysis.
Sec.~\ref{sec:LSS_with_GWs} introduces our fiducial model for astrophysical BHs; throughout this work, unless otherwise specified, we assume a 10-year observational period with a network of three third-generation interferometers, one triangular ET-like and two L-shaped CE-like detectors (ET2CE).
Sec.~\ref{sec:gwbias_with_skao} presents our first analysis: using three independent models for RC, HI-IM, and GW events, we employ a novel version of the publicly available code \texttt{Multi\_CLASS}~\citep{Bellomo:2020pnw, Scarpel_2025} to estimate the constraining power of SKA-Mid$\times$ET2CE on the GW bias.
In Sec.~\ref{sec:gwbias_PBH}, we extend this analysis to estimate the signal-to-noise ratio (SNR) for detecting PBH signatures in the observations.
Going further, Sec.~\ref{sec:analysis_second} explores the impact of different time-delay distributions on GW clustering properties and forecasts the capability of SKAO cross-correlations to constrain the SNR associated with various time-delay models.
Finally, Sec.~\ref{sec:conclusions} summarizes our results and presents our concluding remarks on the promising prospects of using SKAO in cross-correlation analyses for advancing GW science.

%%%%%%%%%%%%%%%%%%%%%%%%%%%%%%%%%%%%%%%%%%%%%%%%%%%%%%%%%%%%%%%%%%%%%%%%%%%%%%%%%%%%%%%%%%%%%%%%%%%%%%%%%%%%%%%%%%

\section{Probing the large-scale structure of the Universe with SKA-Mid surveys}

In this Section, we introduce the formalism needed in our analyses in Sec.~\ref{sec:gwbias_with_skao}, Sec.~\ref{sec:gwbias_PBH} and Sec.~\ref{sec:analysis_second}, to model the cross-correlation between SKAO surveys and another tracer of the LSS.

\label{sec:LSS_with_SKAO}

%%%%%%%%%%%%%%%%%%%%%%%%%%%%%%%%%%%%%%%%%%%%%%%%%%%%%%%%%%%%%%%%%%%%%%%%%%%%%%%%%%%%%%%%%%%%%%%%%%%%%%%%%%%%%%%%%%

\subsection{Brief overview of clustering theory}\label{sec:crosscorr_formalism}

Galaxy clustering has been extensively studied (see, e.g.,~\citep{kaiser, Feldman_1994, Tegmark_1997, Matsubara_1997, Verde_1998, Weinberg_2004, Desjacques_2010, Huterer_2015, DESI:2024mwx}); here we briefly review the theoretical tools used in the analysis presented in this chapter.

Clustering analyses typically focus on deriving the statistical properties of the anisotropies in the fields that describe our tracer of choice, either galaxies, HI-IM or GWs.
Here, we consider statistical properties on large scale, where anisotropic fluctuations are Gaussian and the sky cannot be approximated as a flat surface, and we rely on the two-point function in harmonic space, i.e., the angular power spectrum.
Given two tracers~$X,Y$ in two redshift bins centered at~$z_i,z_j$, respectively, their angular power spectrum reads as~\citep{Challinor:2011bk}
\begin{equation}\label{eq:cl}
    C^{XY}_\ell(z_i,z_j) = 4\pi \int d\log k \ \mathcal{P}_\mathcal{R}(k) \Delta^{X,z_i}_{\ell}(k) \Delta^{Y,z_j}_{\ell}(k)\, ,
\end{equation}
where~$\mathcal{P}_\mathcal{R}$ is the almost scale-invariant primordial curvature power spectrum, and~$k$ and~$\ell$ label Fourier modes and multipoles, respectively.
The harmonic transfer functions are defined as
\begin{equation}\label{eq:transfers}
    \Delta^{X,z_i}_{\ell}(k) = \int_0^\infty dz \frac{dN_X}{dz} W(z, z_i, \Delta z_i) \Delta^X_\ell(k,z)\,,
\end{equation}
where~$dN_X/dz$ describes the redshift-dependent number distribution of the sources,~$W(z,z_i,\Delta z_i)$ is a window function centered at~$z_i$, with half-width~$\Delta z_i$, and normalized to unity.
The full harmonic transfer function~$\Delta^X_\ell(k,z)$ includes density, velocity, and potential terms (see Appendix A of~\cite{Scelfo:2020jyw}); here we highlight the structure of the ``density term'', which is proportional to
\begin{equation}\label{eq:density_term}
    \Delta^{X}_\ell(k,z) \propto \Delta^{X,\mathrm{den}}_\ell (k,z) \propto b_X(k,z) D(k,z),
\end{equation}
where~$D(k,z)$ is the gauge-invariant matter fluctuation, and $b_X(k,z)$ is the scale- and redshift-dependent bias, describing how any given LSS-probe traces the underlying matter distribution~\citep{Sheth_1999, Verde_1998, Sheth_2001, Desjacques:2016bnm}.
In this chapter, $b_X(k,z)=b_X(z)$ always represents the large-scale linear bias, which, in absence of any primordial non-Gaussianity, only depends on time~\citep{Dalal_2008, Matarrese_2008, Desjacques_2010}. Small scales where higher-order bias terms become relevant~\citep{Desjacques:2016bnm} are inaccessible to the poor angular resolution of GW surveys (see Sec.~\ref{sec:LSS_with_GWs}).

In general, the density term in Eq.~\eqref{eq:density_term} is the dominant one for the auto-bin angular power spectrum ($z_i=z_j$). When~$z_i\neq z_j$, instead, the dominant contribution is sourced by ``lensing'' effects~\citep{Yoo_2009, Yoo_2010, Challinor:2011bk, Bonvin11, Jeong_2012, Bertacca_2015_magbias, Raccanelli_2016_2, semenzato2024SFB}.
Additionally, we remind the reader that the angular auto- and cross-power spectra correspond to the cases~$X=Y$ and~$X\neq Y$, respectively, i.e., cross-correlation indicates a scenario where the two tracers are different.

Finally, we note that the angular power spectrum~$C_\ell^{XY}(z_i,z_j)$ described in Eq.~\eqref{eq:cl} only describes the physical signal we plan to observe.
In practice, every measurement suffers from noise, either instrumental, systematic, or intrinsic. For discrete sources like galaxies and GWs, the dominant intrinsic noise is a shot-noise contribution; therefore, the observed angular power spectrum reads as  \begin{equation}
    \tilde{C}_\ell^{XY}(z_i,z_j) = C_\ell^{XY}(z_i,z_j) + \delta^K_{XY} \mathcal{N}^X_\ell\,,
\label{eq:Clnoise}
\end{equation}
where~$\mathcal{N}_\ell$ is the noise angular power spectrum, and the Kronecker delta~$\delta^K$ encodes the fact that noise is typically uncorrelated across different tracers.

%%%%%%%%%%%%%%%%%%%%%%%%%%%%%%%%%%%%%%%%%%%%%%%%%%%%%%%%%%%%%%%%%%%%%%%%%%%%%%%%%%%%%%%%%%%%%%%%%%%%%%%%%%%%%%%%%%

\subsection{SKA-Mid surveys}
\label{sec:SKAO}

SKA-Mid is expected to realize three independent surveys: a neutral-hydrogen (HI) intensity mapping (IM) experiment, a radio-continuum (RC) galaxy survey, and a second, low-redshift, HI-emitting galaxy survey. Because of the low number of detected astrophysical objects and the limited coverage in redshift ($z\in[0.,0.5]$), we do not consider the HI-emitting galaxy survey a good candidate for a cross-correlation study, and we do not include it in our analysis.

All SKAO specifications adopted in this chapter are based on the AA4 `Wide Band 1'' survey described in the SKA Cosmology Red-Book~\citep{SKA_redbook_2020}; further details on the anticipated performance of the SKAO AA* and AA4 designed baseline are described in~\cite{Braun:2019gdo}.
Table~\ref{tab:tracers} summarizes our fiducial survey specifications, together with representative values for the GW survey described in Sec.~\ref{sec:LSS_with_GWs}.
\begin{table}[ht]
	\centerline{
	\begin{tabular}{ccccccc}
		\hline
		Tracer & Experiment &~$z$-range &~$\Delta z$ &~$\ell_{\rm max}$ &~$\Delta\Omega$ [deg$^2$] &~$f_{\rm sky}$\\
		\hline
        \hline
		RC & SKAO &~$[0.0,\ 3.0]$ &~$0.5$ & 500 &~$20\,000\,$ & 0.48\\
		HI-IM & SKAO &~$[0.35,\ 3.0]$ & 0.1 & 200 &~$20\,000$ & 0.48\\
		GW & ET2CE &~$[0.0,\ 3.0]$ & 0.5 & 100,\,200 &~$41\,253$ & 1.0\\
		\hline
	\end{tabular}}
	\caption{Specifics of the surveys in our analysis: redshift range probed; amplitude of each redshift bin~$\Delta z$; multipole associated with the smallest scale resolved~$\ell_{\rm max}$ (for GWs we consider two values, conservative and optimistic respectively, see Sec.~\ref{sec:LSS_with_GWs}); observed sky area~$\Delta\Omega$, and corresponding fraction of the sky~$f_\mathrm{sky}$.}
	\label{tab:tracers}
\end{table}
All tracers (RC, HI-IM, GW) effectively cover the epoch of Cosmic Noon, with HI-IM providing finer redshift sampling.
In terms of angular resolution, the RC survey is able to achieve a much larger angular resolution than the other surveys.
In practice, the maximum multipole~$\ell_{\rm max}$ adopted in this analysis for RC is a conservative limit, especially at high redshift, set by the emergence of nonlinearities in the power spectrum~\citep{Bellomo:2020pnw}.
On the other hand, the maximum multipole is set by instrumental noise and foreground presence for HI-IM, and by intrinsic resolution for the GW interferometer.
The sky coverage of both SKAO surveys is expected to be around~$\Delta\Omega=20\,000\,{\rm deg}^2$ for~$10\,000$ hours of observation, corresponding to an observed fraction of the sky~$f_\mathrm{sky}$ of approximately~$50\%$.
At the same time, a network of GW interferometers is expected to be sensitive to the entire sky, although not in a uniform manner.

%%%%%%%%%%%%%%%%%%%%%%%%%%%%%%%%%%%%%%%%%%%%%%%%%%%%%%%%%%%%%%%%%%%%%%%%%%%%%%%%%%%%%%%%%%%%%%%%%%%%%%%%%%%%%%%%%%

\subsubsection{Radio Continuum Sources}
\label{subsec:RC}

The RC survey in SKA-Mid targets three main categories of sources: star-forming galaxies (SFGs), active galactic nuclei (AGNs), and radio-quiet AGNs.
In SFGs, radio emission has both free-free and synchrotron components, where the former is sourced by ionized HII regions surrounding young stars~\citep{Murphy_2009:ff, Murphy_2013:ff}, while the latter is mostly generated by supernova remnants~\citep{Helou_1985:rc, Condon_1992:rc, Murphy_2006:sr}.
On the other hand, AGNs' emission is sourced by accretion around the central supermassive black hole and by its ejected jets~\citep{Shakura_1972:agn, Begelman_1984:agn, Cattaneo_2009:jet, Best_2012:agn}.
Finally, radio-quiet AGNs represent a sort of intermediate scenario, where radio emission receives contributions both from the SFG and the hosted AGN, see, e.g.,~\cite{Mancuso_2017:rq}.

In this chapter, we rely on catalogs produced using the Tiered Radio Extragalactic Continuum Simulation~(\texttt{T-RECS}, \cite{Bonaldi:2018xfm, Bonaldi_2023}) to model the number density and clustering properties of SFGs and AGNs.
Radio-quiet AGNs are not implemented in \texttt{T-RECS}, and their nature is still under debate; therefore, we do not include them explicitly in our analysis.
However, since their radio emission is dominated by star formation, they would in fact represent a subcategory of the SFG galaxies (see also the discussion in~\cite{Bonaldi:2018xfm}).

We simulated the RC sources baseline catalog by imposing a continuum flux density limit for the AA4 configuration of~$S_{\nu_{\rm c}}=22.8\ \mathrm{\mu Jy}$ at central frequency~$\nu_{\rm c}=797.5\ \mathrm{MHz}$, following the work of~\cite{SKA_redbook_2020}.\footnote{
By design, \texttt{T-RECS} catalogs cover a~$5\times5\,{\rm deg}^2$ area of the sky. The results presented in this chapter have been obtained by rescaling the relations derived to these catalogs to the actual area SKAO will cover, as reported in table~\ref{tab:tracers}.}
The RC sources number per steradian per redshift bin~$d^2N_\mathrm{RC}/dzd\Omega$, which include both SFGs and AGNs, is then fitted with the parametric function
\begin{equation}
    \frac{d^2N_\mathrm{RC}}{dzd\Omega} = \mathcal{A}_\mathrm{RC} \left(\frac{z}{z_\mathrm{RC}}\right)^{\alpha_\mathrm{RC}} e^{-(z/z_\mathrm{RC})^{\beta_\mathrm{RC}}}\,,
\label{eq:dNdzdOmega_RC}
\end{equation}
where the values of best-fit parameters are~$\mathcal{A}_\mathrm{RC}=15516\ \mathrm{gal/deg^2}$, $z_\mathrm{RC} = 0.20$, $\alpha_\mathrm{RC} = 1.38$, and~$\beta_\mathrm{RC} = 0.74$.
On the other hand, the clustering bias of RC sources is estimated using a halo-occupation distribution approach, see, e.g.,~\citep{Cooray_2002:hod, Zheng_2005:hod}.
In this formalism, the average bias is computed for each lightcone slice as a weighted average of individual linear biases, and reads as
\begin{equation}
    b_\mathrm{RC}(z) = \frac{\displaystyle \int_{M^\mathrm{min}_h}^{M^\mathrm{max}_h} dM_h \frac{d^2N_\mathrm{RC}}{dM_hdz} b_h(M_h,z)}{\displaystyle \int_{M^\mathrm{min}_h}^{M^\mathrm{max}_h} dM_h \frac{d^2N_\mathrm{RC}}{dM_hdz}},
\label{eq:bias_RC}
\end{equation}
where~$d^2N_\mathrm{RC}/dM_hdz$ is obtained directly from the~\texttt{T-RECS} catalogs. %, which provide for each source the mass of the host halo~$M_h$.
In the equation, $(M^\mathrm{min}_h,M^\mathrm{max}_h)$ are the minimum and maximum host halo masses in the lightcone slice, and~$b_h(M_h,z)$ is the linear halo bias~\citep{Tinker_2010:bh}.
For later convenience, we fit the bias function through
\begin{equation}
    b_\mathrm{RC}(z) = b^\mathrm{RC}_0 + b^\mathrm{RC}_1 z+ b^\mathrm{RC}_2 z^2+ b^\mathrm{RC}_3 z^3\,,
\label{eq:rc_bias}
\end{equation}
where~$\{ b^\mathrm{RC}_0, b^\mathrm{RC}_1, b^\mathrm{RC}_2, b^\mathrm{RC}_3 \} = \{1.01,-0.19,0.72,-0.08\}$.

\begin{figure}[t]
    \centerline{
    \includegraphics[width=\columnwidth]{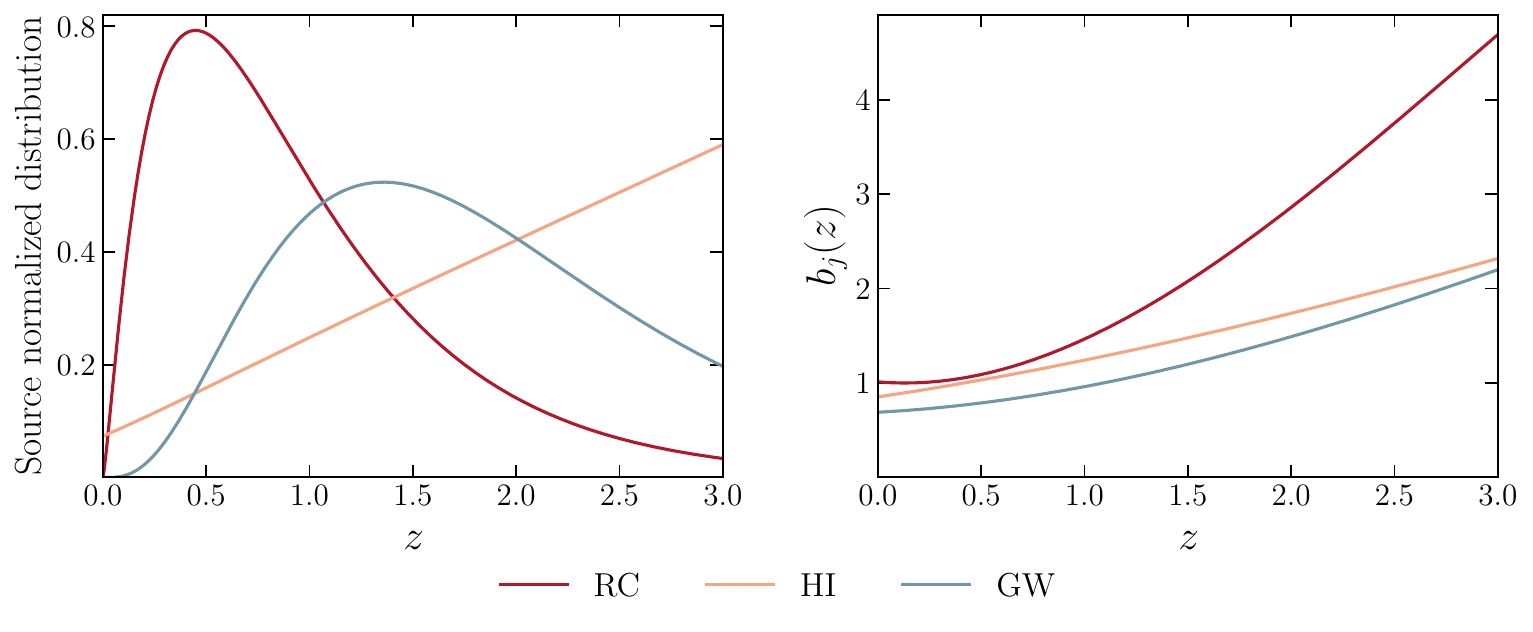}}
    \caption{On the left: normalized redshift-dependent number density distribution of SKAO RC sources (dark red) and ET2CE GW events (blue), and normalized brightness temperature of the HI-IM survey (orange) expected for the Wide Band 1 Survey of SKA-Mid. On the right: redshift-dependent linear bias of the three tracers. Details on the modeling are provided in Sec.~\ref{sec:SKAO} for SKAO tracers, Sec.~\ref{sec:LSS_with_GWs} for GWs (for which we show here the result using \texttt{T-RECS} to link the star formation rate and the host dark matter halo mass).}
\label{fig:SKAO_dNdz_bias}
\end{figure}

Figure~\ref{fig:SKAO_dNdz_bias}, shows the RC number density and bias, in the left and right panels respectively.
As we can observe from the left panel, the bulk of RC is expected to be detected for redshift~$z\lesssim 2$, where star formation peaks around Cosmic Noon.

Finally, for this sample of galaxies, the noise appearing in Eq.~\eqref{eq:Clnoise} has the typical shot-noise form
\begin{equation}
    \mathcal{N}_\ell^\mathrm{RC}(z_i,z_j) = \delta^K_{ij} \left[ \frac{dN_\mathrm{RC}(z_i)}{d\Omega} \right]^{-1} .
\end{equation}
As anticipated in table~\ref{tab:tracers}, in the case of RC survey we set~$\ell^\mathrm{RC}_\mathrm{max}=500$ as maximum resolution, in order to avoid including multipoles that are potentially contaminated by nonlinearities.

%%%%%%%%%%%%%%%%%%%%%%%%%%%%%%%%%%%%%%%%%%%%%%%%%%%%%%%%%%%%%%%%%%%%%%%%%%%%%%%%%%%%%%%%%%%%%%%%%%%%%%%%%%%%%%%%%%

\subsubsection{Neutral Hydrogen Intensity Mapping}
\label{subsec:IM}

At low redshift, neutral hydrogen is mostly found in the interstellar medium of SFGs, making it a perfect candidate target for SKAO telescopes.
The SKA-Mid HI-IM survey will observe the integrated emission of the 21-cm HI line, having rest frame frequency of~$\nu_{\rm rest}\simeq 1420\,{\rm MHz}$, across the entire southern hemisphere of the sky.
This survey will not resolve individual sources, but will measure the (anisotropic) integrated emission of the entire galaxy population, including its faintest and most distant members.
Its observed intensity is usually expressed in terms of the brightness temperature~$T_{\rm HI}$, and thanks to SKA-Mid HI-IM survey outstanding frequency resolution, we will measure it in up to 26 bins between redshift~$z=[0.35,3.0]$.

Although the formalism introduced in this Section has been formally developed for the number count of resolved objects, it can easily be adapted to integrated signals coming from unresolved sources.
Following~\cite{Alonso_2015,Hall_2013,Scelfo:2021fqe}, we change the form of the harmonic transfer functions in Eq.~\eqref{eq:cl} by substituting the number density per redshift with the brightness temperature as
\begin{equation}
    \frac{dN_X}{dz} \longrightarrow T_{\rm HI}(z) = 44\mu {\rm K} \left( \frac{\Omega_{\rm HI}(z)h}{2.45\times 10^{-4}} \right) \frac{H_0(1+z)^2}{H(z)}\,,
\end{equation}
where~$\Omega_{\rm HI}(z)=\rho_{\rm HI}(z)/\rho_{0c}$ is the relative HI density with respect to the critical density~$\rho_{0c}$, $H(z)$ is the Hubble expansion rate, $H_0=H(z=0)$, and~$h=H_0/(100\ \mathrm{km/s/Mpc})$.
The total amplitude is given by the estimates of~\cite{Crighton_2015} and~\cite{Battye_2013}.
As for the HI bias, our model is based on the estimates of~\cite{Spinelli_2020}, which provide an analytical fit of the form
\begin{equation}
    b_{\rm HI}(z) = a_\mathrm{HI}(1+z)^{b_\mathrm{HI}} + c_\mathrm{HI}\,,
\label{eq:hi_bias}
\end{equation}
where~$a_\mathrm{HI}=0.22$, $b_\mathrm{HI}=1.47$, and~$c_\mathrm{HI}=0.63$.
This result is also compatible with the semi-analytical galaxy formation model of~\cite{Villaescusa_Navarro_2018}.
Both the expected brightness temperature and the HI bias are represented in the left and right panels of figure~\ref{fig:SKAO_dNdz_bias}, respectively.

To model the observed angular power spectrum for the HI-IM survey, we consider a instrumental noise characterized by an angular power spectrum of the form~\citep{MeerKLASS:2017vgf, Santos:2015gra}
\begin{equation}
    \mathcal{N}_\ell^{\rm instr}(z_i,z_j) = \delta^K_{ij} \left( \frac{T_{\rm sys}}{T_\mathrm{HI}(z_i) \sqrt{n_{\rm pol} B t_{\rm obs}}} \right)^2 \frac{\Delta\Omega}{N_{\rm d}} \,,
\end{equation}
where, for SKA-Mid,~$T_{\rm sys}=28\,{\rm K}$ is the system temperature, $B=20\,{\rm MHz}$ is the observed frequency bandwidth, $t_{\rm obs}=5\,000\,{\rm hr}$ is the total observation time, $n_{\rm pol}=2$ is the number of observed polarizations, and~$N_{\rm d}=254$ is the number of dishes, as reported in table 2 of~\cite{Santos:2015gra}.

Moreover, since HI-IM collects the integrated emission in the observed band, it also includes contributions from foregrounds, namely astrophysical processes at redshift~$z'<z$ with characteristic emission of photons having rest frame wavelength~$\lambda_{\rm fg}> 21{\rm cm}$, in such a way that~$\lambda_{\rm obs}=(21\,{\rm cm})(1+z)=\lambda_{\rm fg}(1+z')$. The smooth spectral shape of low-$z$ foregrounds allows us to remove their contribution~\citep{Liu:2011hh}, even if not perfectly.
The residual noise due to imperfect foreground cleaning is responsible for noise of the form~\citep{Scelfo:2021fqe}
\begin{equation}\label{eq:noise_fg_lim}
    \mathcal{N}^\mathrm{fg}_\ell (z_i,z_j) = \mathcal{A}_\mathrm{fg} \times \left( \frac{a_\mathrm{fg}}{f_\mathrm{sky}} e^{b_\mathrm{fg}\ell^{c_\mathrm{fg}}} \right).
\end{equation}
where~$\mathcal{A}_\mathrm{fg} = 6\times 10^{-7}$, $a_\mathrm{fg} = 0.129$, $b_\mathrm{fg}=-0.081$ and $c_\mathrm{fg}=0.581$.
These values have been obtained by fitting the scale dependency of the foreground removal found in previous works (see, e.g.,~figure 3 in~\cite{alonso:frgn}). As discussed in~\cite{Scelfo:2021fqe}, the  amplitude~$\mathcal{A}_\mathrm{fg}$ introduces a variance of the residual noise compatible with values in~\cite{camera:frgn}. We note that the residual noise in Eq.~\eqref{eq:noise_fg_lim} is present for both auto- and cross-bin angular power spectra.

Finally, in this analysis, we also consider a small-scale beam-smearing effect due to the single-dish configuration.
Assuming that the beam is Gaussian, we can introduce the beam-smearing factor
\begin{equation}
    \mathcal{B}(z_i) = {\rm exp} \left[ -\ell (\ell+1) \frac{\theta_{\rm IM}^2(z_i)}{{16\log 2}} \right],
\end{equation}
where
\begin{equation}
    \theta_{\rm IM}(z_i) = \frac{1.22 \lambda_{\rm obs}(z_i)}{D_{\rm d}}, \\
\label{eq:angular_resolution}
\end{equation}
is the beam full-width at half-maximum of a single dish with diameter~$D_{\rm d}=15\,{\rm m}$ observing photons with wavelength~$\lambda_{\rm obs}(z)=(21\,{\rm cm})(1+z)$.
Since every spherical harmonic coefficient gets one of these factors, the HI-IM auto- and cross-power spectra signals read as
\begin{equation}\label{eq:cl_obs_im}
    \mathcal{B}(z_i)\mathcal{B}(z_j)C_\ell^{\rm IM,IM}(z_i,z_j), \qquad \mathcal{B}(z_j)C_\ell^{X\neq \rm IM, IM}(z_i,z_j)\,,
\end{equation}
respectively.
Since HI-IM signal at small scales is suppressed by the beam-smearing factor, large multipoles do not contain much information, thus we consider~$\ell^\mathrm{IM}_{\rm max}=200$ across all redshift bins.

%%%%%%%%%%%%%%%%%%%%%%%%%%%%%%%%%%%%%%%%%%%%%%%%%%%%%%%%%%%%%%%%%%%%%%%%%%%%%%%%%%%%%%%%%%%%%%%%%%%%%%%%%%%%%%%%%%

\section{Gravitational waves as tracers of the large-scale structure of the Universe}
\label{sec:LSS_with_GWs}

In the most common astrophysical scenario, BHs form from very massive stars, most likely located in galaxies that underwent an intense star formation epoch.
Consequently, GWs can be used as a tracer of the LSS.
The large number of GW events detected by next-generation interferometers makes this opportunity even more compelling.
 Previous studies of GW clustering have focused on \textit{(i)} inferring the properties of typical host galaxies and different binary formation channels~\citep{Scelfo:2020jyw, Mukherjee:2021bmw, Zazzera:2024agl}, \textit{(ii)} disentangling the presence of exotic GW sources alternative to Astrophysical Black Holes (ABHs)~\citep{Scelfo:2018sny, Calore:2020bpd, Bosi:2023amu, Libanore:2023ovr}, and \textit{(iii)} tightening constraints on cosmological parameters, both in~$\Lambda$CDM and its common extension~\citep{Oguri:2016dgk, Mukherjee:2018ebj, Mukherjee:2020hyn, Mukherjee:2022afz, Libanore:2021jqv, Bosi:2023amu, Gagnon:2023mnd, Afroz_2024, Balaudo:2023klo, Pedrotti:2025tfg}.
Although the majority of cross-correlation studies focused on resolved GW events in the Hz-kHz frequency band, the technique has also been used to study cross-correlations between LSS and the GW background in the nHz band~\citep{sah_2024,Semenzato:2024mtn,Sah:2025uuk}.
The interested reader can find a more in-depth discussion regarding this latter point in~\cite{Ragavendra01.2026.SKA}.
In this chapter, we focus on the first two scientific goals of the above list, which are closely linked with the physics of GW sources.

As with galaxy number counts and neutral hydrogen brightness temperature,  the formalism of Sec.~\ref{sec:crosscorr_formalism} applies directly to GW number counts.
Previous works suggest that the angular power spectrum in Eq.~\eqref{eq:cl} is more naturally defined in luminosity distance space, as its measurement from GW data does not rely on any cosmological model to transform distances into redshifts ~\citep{Zhang:2018nea, Namikawa:2015prh, Namikawa:2020twf, Libanore:2020fim, Fonseca:2023uay}. Despite this, the uncertainty in the luminosity distance measurements in GW data is much larger compared to the uncertainty on cosmological parameters, and hence dominates the uncertainty budget. The good constraints from {\it Planck 2018}~\citep{Planck:2018vyg}, expected to be further improved by current and upcoming CMB and LSS surveys, justifies the carrying out in redshift-space of GW$\times$galaxy analyses as the one in this chapter, expected to be realized in the mid 2030s.

 Our fiducial scenario assumes all GW events are sourced by astrophysical black hole (ABH) mergers, following~\cite{Bosi:2023amu,bellomo:classgwb}.
 Here, the binary formation rate is proportional to the cosmic star-formation rate (SFR), the time-delay distribution follows~$p(t_d)\propto t_d^{-1}$, and the local merger rate matches the LIGO-Virgo-KAGRA (LVK) estimate,~$\mathcal{R}^{\rm LVK}_0 \simeq 20\,{\rm Gpc}^{-3}{\rm yr}^{-1}$~\citep{KAGRA:2021duu}.
 For a third-generation (ET2CE), composed by one triangular ET-like and two L-shaped CE-like detectors, the GW number density per redshift and steradian is parametrized as
\begin{equation}\label{eq:dnBH}
    \frac{d^2N_{\rm GW}}{dzd\Omega} = \mathcal{A}_\mathrm{GW} \left(\frac{z}{z_\mathrm{GW}}\right)^{\alpha_\mathrm{GW}} e^{-(z/z_\mathrm{GW})^{\beta_\mathrm{GW}}}\,,
\end{equation}
where~$\mathcal{A}_\mathrm{GW} = 1.35\times 10^{-7}\ \mathrm{GW/sr}$, $z_\mathrm{GW} = 2.02\times 10^{-3}$, $\alpha_\mathrm{GW}=6.12$, and~$\beta_\mathrm{GW}=0.41$.

We model the GW bias as
\begin{equation}
    b^\mathrm{TRECS}_{\rm GW}(z) = b^\mathrm{GW}_0 + b^\mathrm{GW}_1 z + b^\mathrm{GW}_2 z^2 + b^\mathrm{GW}_3 z^3,\,
\label{eq:gw_trecs_bias}
\end{equation}
where the parameters~$b^\mathrm{GW}_0=0.68$, $b^\mathrm{GW}_1=0.12$, $b^\mathrm{GW}_2=0.17$, $b^\mathrm{GW}_3=- 0.01$, are fitted to match the \texttt{T-RECS} catalog (see~\cite{bellomo:classgwb} for the methodology). In Appendix~\ref{app:UM} we test the robustness of our model when adopting a different parametrization gauged on the results of \textsc{UniverseMachine} (UM), see e.g.~\cite{behroozi:universemachine}.

Since GWs are discrete tracers, similarly to galaxy surveys, the measurement of the GW clustering signal is contaminated by shot-noise, with an angular power spectrum of the form~\citep{Bosi:2023amu}
\begin{equation}
    \mathcal{N}^\mathrm{GW}_\ell (z_i,z_j) = \delta^K_{ij} \left[ \frac{dN^\mathrm{obs}_\mathrm{GW}(z_i)}{d\Omega} \right]^{-1} = \delta^K_{ij} \left[ \int dz \frac{d^2N_\mathrm{GW}}{dz d\Omega} p_\mathrm{obs}(z_i,z) \right]^{-1},
\end{equation}
where~$p_\mathrm{obs}(z_i,z)$ is the probability that a GW emitted at redshift~$z$ is observed in the $z_i$ bin, accounting for localization uncertainty.
We choose redshift bins that are wider than the typical GW localization error, hence the specific form of~$p_\mathrm{obs}(z_i,z)$ has negligible impact on our results.

Finally, we remind the reader that the precise and accurate sky localization of a GW event requires a network of multiple detectors to perform triangulation.
However, even in the case of ET2CE, the precision of the sky localization ranges between~$[10^{-3}, 10^2]\,{\rm deg}^{2}$~\citep{Iacovelli:2022bbs}, with an average precision of a few square degrees~\citep{Calore:2020bpd, Bosi:2023amu}, depending on the binary and network properties at the moment the signal arrives on Earth. This prevents GW measurements to resolve small scales,~i.e.,~large multiples $\ell$s in the angular power spectrum in Eq.~\eqref{eq:angular_resolution}. As anticipated in Table~\ref{tab:tracers}, we consider two scenarios: a conservative scenario with~$\ell^\mathrm{GW}_{\rm max}=100$, as in~\cite{Scelfo:2021fqe}, and an optimistic one with~$\ell^\mathrm{GW}_{\rm max}=200$, as in~\cite{Bosi:2023amu}.

%%%%%%%%%%%%%%%%%%%%%%%%%%%%%%%%%%%%%%%%%%%%%%%%%%%%%%%%%%%%%%%%%%%%%%%%%%%%%%%%%%%%%%%%%%%%%%%%%%%%%%%%%%%%%%%%%%

\section{Improving the measurement of GW bias with SKAO}
\label{sec:gwbias_with_skao}

Our first goal is to forecast the sensitivity on GW bias parameters obtained using the GW dataset alone, and to contrast it with the case where SKA-Mid$\times$ET2CE cross-correlation is implemented.
We begin our analysis assuming that the GW survey contains exclusively ABHs.
For the rest of this work, we fix the cosmological parameters to their \textit{Planck 2018} TT,TE,EE+lowE+lensing best-fit values~\citep{Planck:2018vyg}.
The angular auto- and cross-power spectra are computed with the updated version of \texttt{Multi\_CLASS}~\citep{Bellomo:2020pnw, Scarpel_2025} for all tracers described in Sec.~\ref{sec:LSS_with_SKAO} and Sec.~\ref{sec:LSS_with_GWs}, using as fiducial models the ones previously described.
We consider a different number of redshift bins for each tracer,~$N_{\rm bins}^{\rm RC}=6$, $N_{\rm bins}^{\rm IM}=26$, and~$N^{\rm GW}_\mathrm{bins}=6$.
Each redshift bin is characterized by a top-hat window function with half-widths~$\Delta z_{\rm RC}=0.5$, $\Delta z_{\rm IM}=0.1$ (except for the last bin that has~$\Delta z = 0.15$), and~$\Delta z_{\rm GW}=0.5$, as anticipated in table~\ref{tab:tracers}.

We perform a Fisher matrix analysis~\citep{fisher:fisher, bunn:fisher, vogeley:fisher, tegmark:fisher} to estimate the future sensitivity of GW datasets alone and in cross-correlation with either the RC or HI-IM surveys of SKAO.
The set of parameters analyzed in each scenario is:
\begin{equation}
    \begin{aligned}
        \bm{\theta}_\mathrm{GW} &= \left[ b_{\rm GW,1}, ..., b_{\rm GW,6} \right]\,, \\
        \bm{\theta}_\mathrm{RC\times GW} &= \left[ n_s, \omega_\mathrm{cdm}, h, b_{\rm GW,1}, ..., b_{\rm GW,6}, b_{\rm RC, 1}, ..., b_{\rm RC,6} \right]\,, \\
        \bm{\theta}_\mathrm{HI-IM\times GW} &= \left[ n_s, \omega_\mathrm{cdm}, h, b_{\rm GW,1}, ..., b_{\rm GW,6}, b_{\rm IM, 1}, ..., b_{\rm IM,26} \right]\,, \\
    \end{aligned}
\end{equation}
where~$n_s$ is the spectral index of scalar perturbations, $\omega_\mathrm{cdm}$ is the cold dark matter physical density, $h=H_0/(100\ \mathrm{km/s/Mpc})$, and the fiducial values of the bias parameters~$b_{ X,j}$ for the tracer~$X=\mathrm{GW,RC,IM}$ in the $j$-th redshift bin are obtained from Eq.~\eqref{eq:rc_bias} for RC, Eq.~\eqref{eq:hi_bias} for HI-IM, and either Eq.~\eqref{eq:gw_um_bias} or Eq.~\eqref{eq:gw_trecs_bias} for GWs.
Additionally, for LSS surveys with angular resolution greater than the GW one, we use the additional information contained at multipoles~$\ell > \ell^\mathrm{GW}_\mathrm{max}$ as detailed in~\cite{Bellomo:2020pnw}.

\begin{figure}[ht]
    \centerline{
    \includegraphics[width=\columnwidth]{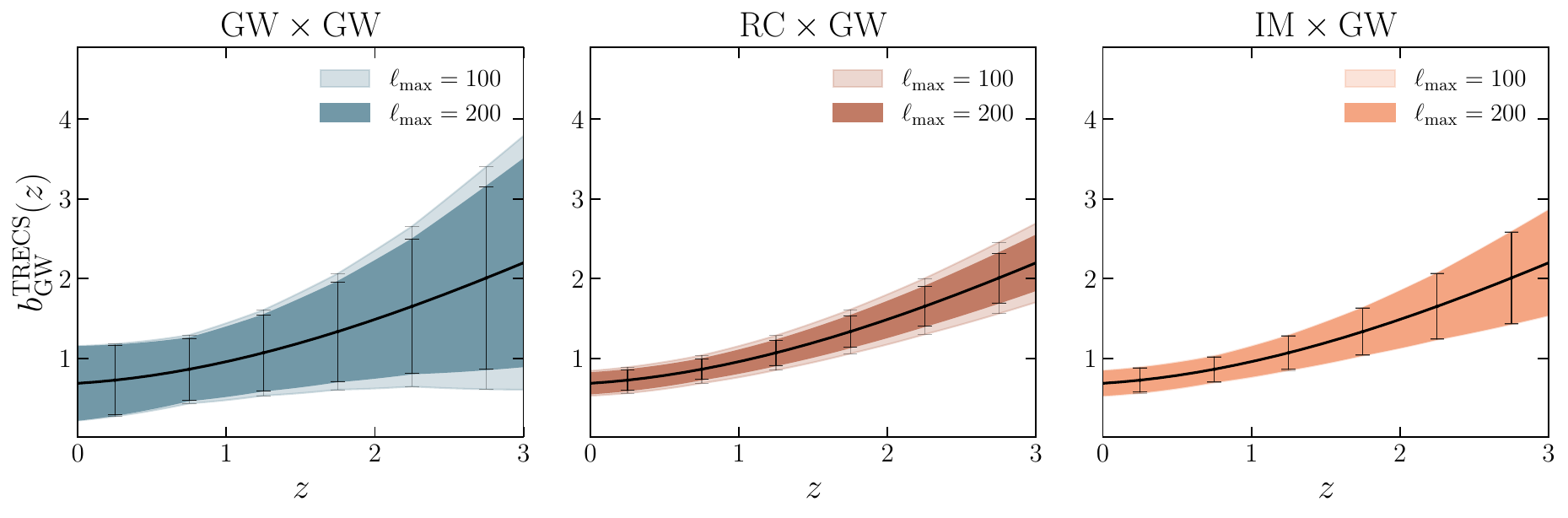}}
    \caption{Forecast constraints on the GW bias using ET2CE alone (\textit{left panel}), in cross-correlation with RC galaxies (\textit{middle panel}), and with the HI-IM (\textit{right panel}) survey observed by SKA-Mid. The errorbars show the forecasts per each $z$-bin, while the shaded area extrapolates to the full redshift range. Each panel shows results for two choices of $\ell_{\rm max}$ for the GW survey.
    }
\label{fig:sigma_bGW_ST_MT}
\end{figure}

Our result is shown in figure~\ref{fig:sigma_bGW_ST_MT}, where we marginalize over the cosmological parameters and, for scenarios involving either RC or HI-IM surveys, over the bias of the SKAO tracer. The constraining power of the cross-correlation largely overcomes forecasts for ET2CE alone, demonstrating once more the importance of this kind of study to foster our understanding of GW clustering.
Since in all cases parameters describing both the GW bias and the bias of the SKAO-tracer present partial degeneracies with the cosmological parameters, the constraints can be further improved by imposing a prior on the cosmological model, from,~e.g.,~{\it Planck 2018}.

Our results for RC galaxies and HI-IM are comparable. On the one hand, HI-IM tomographic precision allows us to better probe the redshift evolution of the bias. On the other, the larger parameter set due to the~$b_{\rm IM}^i$ parameters makes the Fisher more sensitive to degeneracies among their values and the~$b_{\rm GW}^i$ parameters, thus lowering the constraining power.
Also, the RC survey benefits from the contribution of more signal coming from  small scales in the auto-power spectra, which is damped in the HI-IM case due to the single-dish beam.
Furthermore, we test two different choices of $\ell_{\rm max}$ for the GW survey, as reported in table~\ref{tab:tracers}. The more optimistic value, $\ell_{\rm max}=200$, yields only a modest improvement in constraining power. This is because, in our fiducial setup, the GW auto-power spectra are already shot-noise dominated at $\ell\sim100$. Including smaller scales therefore adds little information, especially in the HI-IM case, where beam smearing further suppresses small-scale sensitivity. Although futuristic, assuming a longer GW observational time would reduce the impact of shot noise, making small-scale information more relevant.

To conclude, we note that the GW bias constraints may vary when changing the ABH, RC or HI-IM bias models. As discussed in Appendix~\ref{app:UM}, changes in the calibration of the GW bias in Eq.~\eqref{eq:gw_trecs_bias} have a small impact on the final results; on the other hand, a different redshift evolution can have a stronger effect. According to the authors of~\cite{Bosi:2023amu} and~\cite{Zazzera_2025:ska_cc}, the cross correlation with SKAO may lead to more stringent forecasts than the ones described in this chapter: instead of using the full set of $b_{\rm RC}^{i_1}$,~$b_{\rm IM}^{i_2}$,~$b_{\rm GW}^j$ parameters, they rely either on the effective bias (i.e., a single parameter obtained by averaging over all the redshift bins), or on a power-law parametrization of the bias, deriving constraints on its parameters. Despite being less optimistic, our analysis allows for more degrees of freedom, which we believe are important considering that prior knowledge of the bias of these tracers is still poor.

%%%%%%%%%%%%%%%%%%%%%%%%%%%%%%%%%%%%%%%%%%%%%%%%%%%%%%%%%%%%%%%%%%%%%%%%%%%%%%%%%%%%%%%%%%%%%%%%%%%%%%%%%%%%%%%%%%

\section{Constraining the presence of exotic GW sources}\label{sec:gwbias_PBH}

Astrophysical BHs are not the unique GW-emitting candidate compatible with current observations.
A second alternative, which has raised considerable attention over the past decade, is given by primordial black holes (PBHs), which may constitute a fraction $f_{\rm PBH}$ of dark matter~\citep{zeldovich:pbhformation,hawking:pbhformation,carr:pbhformation,chapline:pbhformation,sasaki:pbhconstraintsreview}.
Depending on the PBH mass, $f_{\rm PBH}$ is constrained by a variety of astrophysical and cosmological observables, see, e.g.,~\citep{Carr:2020xqk, Carr:2020gox} for recent reviews; however, the robustness of these bounds is still a matter for debate, as shown, e.g., in~\cite{Piga_2022}.

PBHs with masses comparable to those of ABHs can bound in binaries in both the early (EPBH,~\cite{Ali-Haimoud:2017rtz, Raidal:2018bbj}) and late (LPBH,~\cite{Bird:2016dcv,Clesse:2016vqa}) Universe, eventually merging and emitting GWs.
Although the mass spectrum of these compact objects can resemble that of astrophysical sources, their redshift distribution and clustering properties is radically different.

Given the profound impact that the discovery of PBHs would have, in this work we also discuss the possibility of disentangling the presence of PBHs in the midst of an ABHs sample.
While in our fiducial scenario we model the GW number density and bias assuming that we only have ABHs, now we also consider the possibility of having contributions from EPBHs and LPBHs.

\begin{figure}[ht]
    \centerline{
    \includegraphics[width=1\columnwidth]{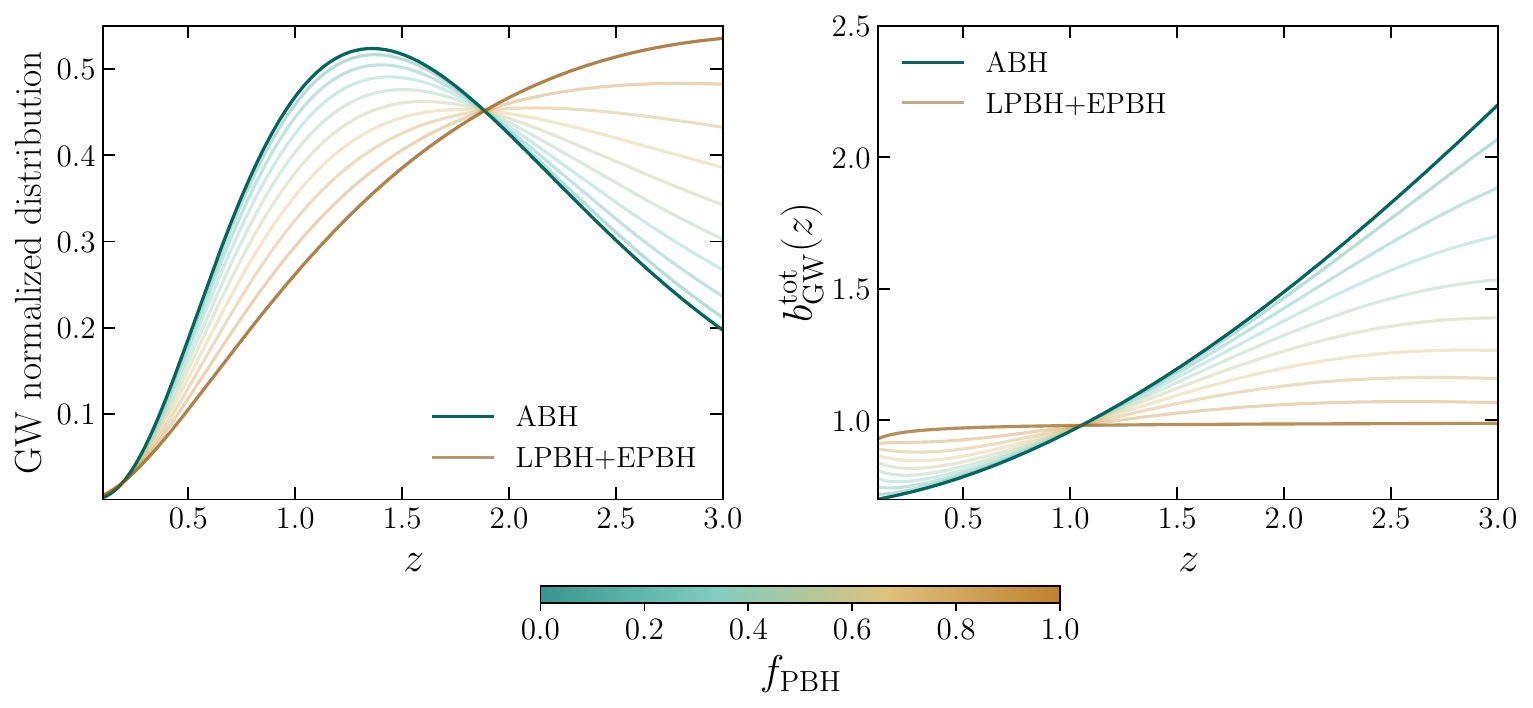}}\vspace*{-.3cm}
    \caption{Redshift distribution (left) and bias (right) of GW events expected for ET2CE.
    In the~$f_{\rm PBH}=0$ scenario (blue), only ABHs are present; when~$f_{\rm PBH}=1$ (brown) the survey only contains PBHs.
    The color-coded curves show intermediate scenarios, with relative ABHs, EPBHs, LPBHs abundances set by~$f_{\rm PBH}$. }
\label{fig:ABH_PBH_dNdz_bias}
\end{figure}

We report in figure~\ref{fig:ABH_PBH_dNdz_bias} the normalized number density distribution~$d^2N_{\rm GW}/dzd\Omega$ and the GW bias~$b_{\rm GW}$ in scenarios where the relative importance of the ABH and EPBH+LPBH contributions varies according to~$f_{\rm PBH}$. The ABH model follows the description in Sec.~\ref{sec:LSS_with_GWs}, while the EPBH+LPBH contribution is obtained using the same formalism as in~\cite{Bosi:2023amu}, with a fiducial EPBHs merger rate of~$\mathcal{A}_\mathrm{m}=18\ \mathrm{Gpc^{-3}yr^{-1}}$.
The blue line corresponds to the scenario in which only ABHs are present, i.e., $f_{\rm PBH}=0$.
In contrast, the brown curve refers to the case with only PBHs ($f_{\rm PBH}=1$).
In between these two scenarios, the color-coded curves describe mixed cases, including ABHs, EPBHs and LPBHs, depending on the value of~$f_{\rm PBH}$.

Finally, we forecast the ability of SKA-Mid$\times$ET2CE cross correlation to detect the contribution of LPBHs and EPBHs. We quantify this through the signal-to-noise ratio (SNR) of the difference between biases in the fiducial ABH case and alternative ABH-PBH mixed scenario, defined as
\begin{equation}
    \begin{aligned}
        {\rm SNR}^2(f_{\rm PBH},\mathcal{A}_m) &= \sum_j {\rm SNR}^2 (f_{\rm PBH}, \mathcal{A}_m, z_j) = \sum_j \frac{\left( b^\mathrm{alt}_{\rm GW} (f_{\rm PBH}, \mathcal{A}_m, z_j) - b^\mathrm{fid}_{\rm GW}(z_j) \right)^2}{\sigma^2_{b^\mathrm{fid}_{\rm GW}}(z_j)}\,,
    \end{aligned}
\end{equation}
where we also allowed for variations in the EBH local merger rate, $\mathcal{A}_m$, to account for uncertainties in their abundance~\citep{Raidal:2018bbj, Ballesteros:2018swv, DeLuca:2020jug}. In the equation, the values of~$\sigma^2_{{b}_{\rm GW}^\mathrm{fid}}$ are those obtained in Sec.~\ref{sec:gwbias_with_skao}.

\begin{figure}[ht]
    \centerline{
    \includegraphics[width=\linewidth]{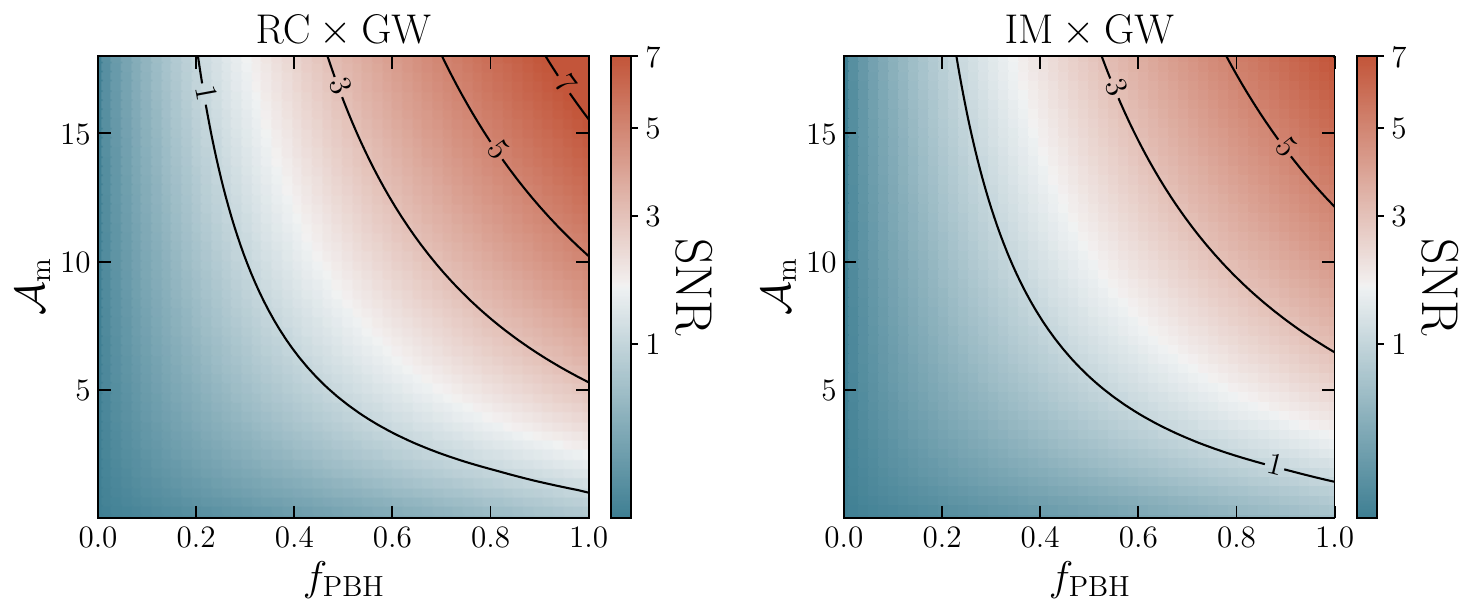}}
    \caption{Total SNR of the difference between only-ABH and ABH+EPBH+LPBH scenarios, varying~$\mathcal{A}_m$,~$f_{\rm PBH}$.
    On the left, we cross-correlate GW with RC, on the right with HI-IM.}
\label{fig:pbh_summary}
\end{figure}

Figure~\ref{fig:pbh_summary} shows the total SNR obtained by summing over all the redshift bins.
Our forecasts show that %for values of~$\mathcal{A}_m$ close to the fiducial value expected from theoretical studies,
~$f_{\rm PBH}\in [0.8,1]$ will be constrained by SKA-Mid$\times$ET2CE at~$>5\sigma$ level; smaller fractions will be constrained with a smaller confidence level.
Overall, RC cross correlation will have larger constraining power than HI-IM. However, due to uncertainties in the EPBH formation mechanism~\citep{Raidal:2018bbj, Ballesteros:2018swv, DeLuca:2020jug}, a smaller amplitude $\mathcal{A}_m$ may lead to constraints with a smaller confidence level (see~\cite{Bosi:2023amu}).

Our results are in a similar ballpark of current constraints on~$f_{\rm PBH}$ in the PBH mass range comparable to ABHs probed by GW observations~$M_{\rm PBH}\in[1\,M_\odot,100\,M_\odot]$ (compare, e.g., with Figure 10 in~\cite{Carr:2020gox}), confirming that the cross-correlation between SKA-Mid and ET2CE will strengthen our interpretation of GW data, confirming or ruling out the existence of PBH binaries.

%%%%%%%%%%%%%%%%%%%%%%%%%%%%%%%%%%%%%%%%%%%%%%%%%%%%%%%%%%%%%%%%%%%%%%%%%%%%%%%%%%%%%%%%%%%%%%%%%%%%%%%%%%%%%%%%%%

\section{Constraining the time-delay distribution}
\label{sec:analysis_second}

Phenomenological models such as those presented in Sec.~\ref{sec:LSS_with_GWs} are widely used to connect the observed merger rate to the underlying physical properties of GW sources and hosts, especially with the SFR, although metallicity has also been shown to play an important role~\citep{Langer:2005hu, Ma:2015nya, Chruslinska:2018hrb}.
However, drawing this connection is not straightforward, and even in the ABH scenario our understanding of GW progenitors remains limited.
Although SFR and metallicity set the BBH formation rate, the time required for a binary to merge also depends on the time-delay distribution, which is primarily determined by the BBH formation mechanism and the local environment.
Short time-delays are typically associated with environments with low metallicity~\citep{Marchant:2016wow, duBuisson:2020asn}, and binaries formed inside young star clusters~\citep{DiCarlo:2020lfa} or in AGN disks~\citep{Yang:2020lhq}.
Mergers occurring within globular clusters are also characterized by short time-delays, unless the BBH is ejected from the cluster~\citep{Benacquista:2011kv, Rodriguez:2016kxx, Banerjee:2016ths, Rodriguez:2017pec}.
In contrast, isolated binary evolution predicts a statistical distribution of time-delays scaling as~$p(t_d)\propto t_d^{\alpha_d}$, where the typical value~$\alpha_d = -1$ follows from the assumption that the distribution of BBH semi-major axis at formation, $a$, scales as~$p(a) \propto a^{-1}$~\citep{Abt:1983tr}.
However, the validity of this assumption is still uncertain and it depends very likely on additional effects such as stellar winds, mass transfer, and natal kicks~\citep{OShaughnessy:2007brt, OShaughnessy:2009szr, Mapelli:2017hqk}.

Constraining the time-delay distribution from GW-only experiments represents a challenging endeavor, since the observed merger rate probes a degenerate combination of the time-delay distribution and the binary formation rate.
However, GW clustering (in this case in cross-correlation with SKAO LSS surveys) provides an additional handle to constraint the effective time-delay distribution because it breaks this intrinsic degeneracy.
Here we report the preliminary results discussed in~\cite{Bellomo_2026_prep}, where two cases are discussed: a first scenario where we fix the binary formation rate and we vary the time-delay distribution, and a second one where we fix the merger rate while varying the time-delay distribution.
In all the cases the local merger rate is compatible with the observed Ligo-Virgo-KAGRA value, as explained in Sec.~\ref{sec:LSS_with_GWs}.

\begin{figure}[ht]
    \centerline{
    \includegraphics[width=\linewidth]{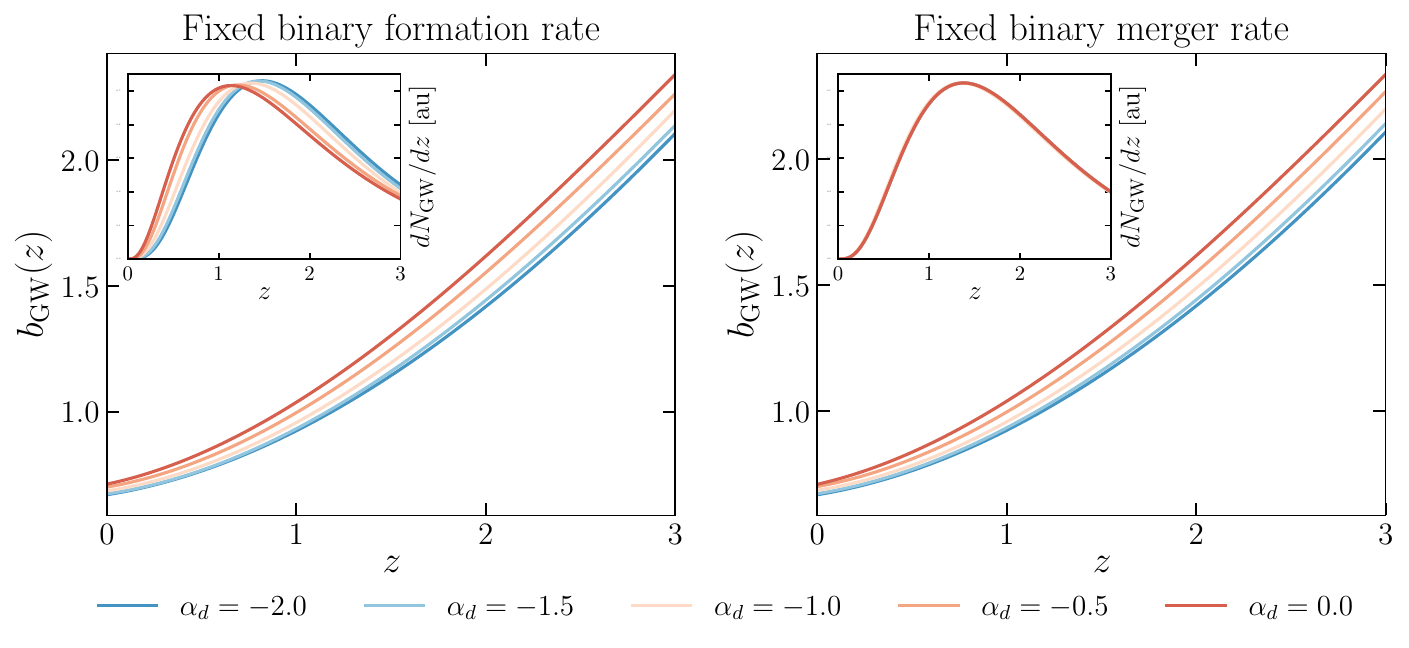}}
    \caption{GW bias obtained for different choices of the time-delay power-law distribution, and fixing either the BBH formation rate (\textit{left panel}) or their merger rate (\textit{right panel}).
    \textit{Insets:} GW redshift-dependent number density distribution in arbitrary units (au) for the same two scenarios, following the same conventions of the main panels. In both cases, we model the SFR-$M_h$ relation using the \texttt{T-RECS} parameterization.
    }
    \label{fig:dNdz_bias_timedelay}
\end{figure}

The GW number density and bias corresponding to the two scenarios are reported in Fig.~\ref{fig:dNdz_bias_timedelay}.
Different curves correspond to different time-delay distributions parameterized as~$p(t_d) \propto t_d^{\alpha_d}$, where~$\alpha_d=-1$ for our fiducial model.
In the first case, the merger rate depends on the time-delay choice, which is responsible for a different shape of number density distribution.
On the other hand, in the second scenario, the merger rate is constant across models, since what changes is the binary formation rate.
In both cases, the GW bias depends on the choice of time-delay distribution and, most notably, it still presents some variability even in the fixed merger rate scenario, where number densities of different models are fundamentally indistinguishable.
In other words, information about the time-delay distribution is intrinsically encoded in the clustering properties of GWs.

We quantify the statistical difference in clustering properties associated with different time-delay distributions by introducing a signal-to-noise ratio defined as
\begin{equation}
    \chi^2 = f_{\mathrm{sky}} \sum_{\ell} \frac{2\ell+1}{2} \mathrm{Tr} \left[ \Delta\mathcal{C}_\ell \left( \tilde{\mathcal{C}}^\mathrm{fid}_{\ell} \right)^{-1} \Delta\mathcal{C}_\ell \left( \tilde{\mathcal{C}}^\mathrm{fid}_{\ell} \right)^{-1} \right],
\label{eq:chi2statistics}
\end{equation}
where~$\Delta\mathcal{C}_\ell = \mathcal{C}^\mathrm{alt}_\ell - \mathcal{C}^\mathrm{fid}_{\ell}$ is the difference between the covariance matrices of the alternative and fiducial time-delay models, while~$\tilde{\mathcal{C}}^\mathrm{fid}_{\ell}$ is the covariance matrix of the fiducial model.

\begin{figure}[ht]
    \centerline{
    \includegraphics[width=\linewidth]{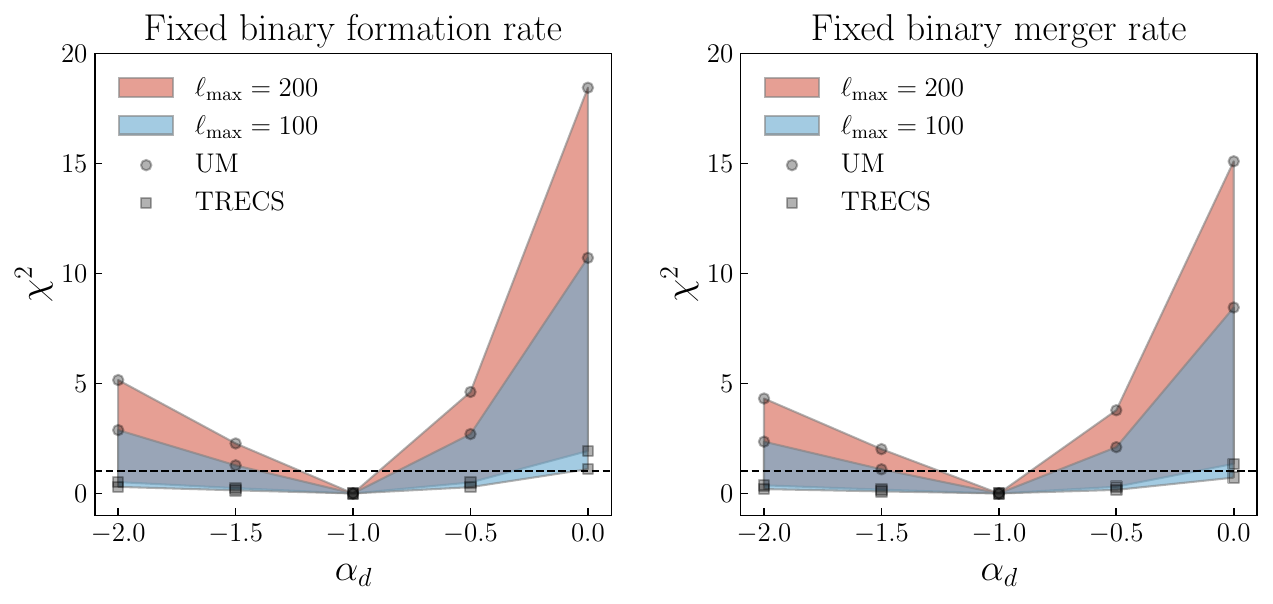}}
    \caption{$\chi^2$ detectability of different time-delay distributions~$p(t_d)\propto t_d^{\alpha{_d}}$, with respect to the fiducial case~$\alpha_d=-1$.
    Results are obtained from the RC$\times$GW cross-angular power spectra, by modeling the GW bias as function of~$p(t_d)$ and of the SFR of the host galaxies, with the \texttt{T-RECS} ({\it square markers}) and UM ({\it circle markers}) parametrization.
    More conservative results ($\ell_{\rm max}=100$) are shown in blue, more optimistic ($\ell_{\rm max}=200$) in red.
    We test both what happens when the normalization is set fixing the formation rate ({\it left panel}) or the merger rate ({\it right panel}).}
\label{fig:chi2statistics_GWRCSKA_MT}
\end{figure}

The results of this kind of analysis are reported in Fig.~\ref{fig:chi2statistics_GWRCSKA_MT}, which shows the statistical distinguishability between different time-delay models.
The bias characterizing a GW population is the average of the halo bias weighted by the probability~$p(M_h)$ of finding a GW in host halo with a given mass.
Since different time-delay distributions change the association between GW events and host halo masses, clustering measurements have the potential to distinguish between different scenarios.
In fact, the quantitative differences in~$\chi^2$ originate from their different parameterizations of the SFR–$M_h$ relation, leading to a different~$p(M_h)$ at binary formation.
Our forecasts show that the imprint of the time-delay distribution on the GW clustering is robustly detectable.
In other words, the cross-correlation of GW clustering measurements with SKAO LSS surveys provides a novel powerful observable to probe the BBH formation mechanism.

%%%%%%%%%%%%%%%%%%%%%%%%%%%%%%%%%%%%%%%%%%%%%%%%%%%%%%%%%
%%%%%%%%%%%%%%%%%%%%%%%%%%%%%%%%%%%%%%%%%%%%%%%%%%%%%%%%%
%%%%%%%%%%%%%%%%%%%%%%%%%%%%%%%%%%%%%%%%%%%%%%%%%%%%%%%%%

\section{Conclusions}
\label{sec:conclusions}

Binary black hole mergers trace the Large-Scale Structure of the Universe (LSS), but extracting their clustering properties from Gravitational Wave (GW) observations faces fundamental challenges.
Next-generation detectors such as the Einstein Telescope and the Cosmic Explorer will detect millions of events over the coming decades; yet, poor localization of events in terms of distance and sky localization constitute an intrinsic limit in their ability to map the cosmic web.
Cross-correlating GW catalogs with LSS surveys offers a direct strategy to mitigate both instrumental and theoretical uncertainties that hinder our understanding of GW sources and properties.

The SKAO, with its SKA-Mid neutral hydrogen intensity mapping (HI-IM) and radio continuum (RC) surveys, provides the ideal datasets for a synergistic cross-correlation analysis with GWs.
Wide sky coverage and deep redshift sensitivity extending up to~$z=3$ overlap precisely with the epoch when astrophysical Black Hole (BH) binaries are expected to form most efficiently.
In particular, the RC survey has the potential of detecting a large fraction of Star Forming Galaxies (SFGs), providing an accurate map of the most-likely environments where ABH binaries originate.

In this chapter, we have demonstrated the scientific potential for the SKAO surveys when used in cross-correlation with GW datasets coming from third-generation GW detector networks.
Independently of the chosen SKAO LSS survey, error bars on the reconstructed bias parameters shrink by a factor of few when compared to the GW-only analysis, with some additional gain in constraining power for the RC survey  due to its large number densities and access to small scales.
Because of this unprecedented sensitivity, cross-correlation studies have the potential to disentangle the presence of different BH populations if they are characterized by different clustering properties.
In particular, we focused our interest on the possibility of detecting the presence of PBHs.
Our analysis shows that SKA-Mid$\times$ET2CE achieves~$>5\sigma$ detection for primordial black hole fractions~$f_{\rm PBH} \geq 0.8$ at theoretically motivated merger rate amplitudes~$\mathcal{A}_m \sim 18\,{\rm Gpc}^{-3}{\rm yr}^{-1}$.
This detection relies entirely on differences in clustering bias between astrophysical and primordial populations, since GW observations can hardly determine the nature of BHs.

Aside from the presence of exotic GW-emitting sources, we also showed how the cross-correlations with SKAO address the degeneracy between the binary formation rate and the time-delay distribution between formation and merger.
The observed GW merger rate is a convolution of these two quantities, making it exceptionally difficult to disentangle the underlying physics of binary BH formation channels with GW data alone.
Even when the observed merger rate is fixed (thus maintaining identical GW redshift distributions), different time-delay models produce distinct clustering signatures, leaving a unique, detectable imprint on the clustering bias of the GW sources.
In particular, our analysis shows how it is possible to effectively measure the time-delay parameters distribution with large statistical significance.
This result opens a new avenue for constraining the dominant binary BH formation mechanisms, a crucial open problem in GW astrophysics.

Looking ahead, the synergy between third-generation GW observatories and SKAO heralds a new era of precision multi-messenger cosmology and astrophysics.
In this work, we have focused exclusively on binary BH mergers, as they will constitute the bulk of detections at the large cosmological distances probed by these instruments.
However, the intrinsic merger rates of binary neutron star (BNS) and neutron star-black hole (NS-BH) systems are expected to be substantial.
Once detectors will achieve the sensitivity to build large catalogs of these events, our cross-correlation framework will readily extended to them.
This will provide independent constraints on their distinct formation channels and time-delay distributions.
SKA1-Mid extensions to higher frequencies will detect AGN and starburst galaxies with exquisite precision, providing complementary tracers of the environments hosting dynamical formation channels.
Ultimately, by leveraging the statistical power of LSS surveys such as SKAO to calibrate the astrophysics of GW sources, we can turn these messengers into cleaner probes of fundamental physics and cosmic expansion history.

%%%%%%%%%%%%%%%%%%%%%%%%%%%%%%%%%%%%%%%%%%%%%%%%%%%%%%%%%%%%%%%%%%%%%%%%%%%%%%%%%%%%%%%%%%%%%%%%%%%%%%%%%%%%%%%%%%

\appendix
\section{Alternative GW parametrization}\label{app:UM}

To model the GW bias in the main text we relied on the method described in~\cite{bellomo:classgwb}, and calibrated it to match the~\texttt{T-RECS} simulations. In the original work, the bias was calibrated on \textsc{UniverseMachine} (UM), which differs from \texttt{T-RECS} in the way it connects the star-formation rate to dark matter halo mass. In this Appendix, we test whether our results are robust to this model change.
By implementing the~UM parametrization, we find that the GW bias is well-described by the fitting formula
\begin{equation}
    b^\mathrm{UM}_{\rm GW}(z) = b^\mathrm{GW}_0 + b^\mathrm{GW}_1 z + b^\mathrm{GW}_2 z^2 + b^\mathrm{GW}_3 z^3,\,
\label{eq:gw_um_bias}
\end{equation}
where the best-fit parameters are given by~$b^\mathrm{GW}_0=0.65$, $b^\mathrm{GW}_1=1.45$, $b^\mathrm{GW}_2=-0.14$, $b^\mathrm{GW}_3=0.01$.

\begin{figure}[ht]
    \centerline{
    \includegraphics[width=\columnwidth]{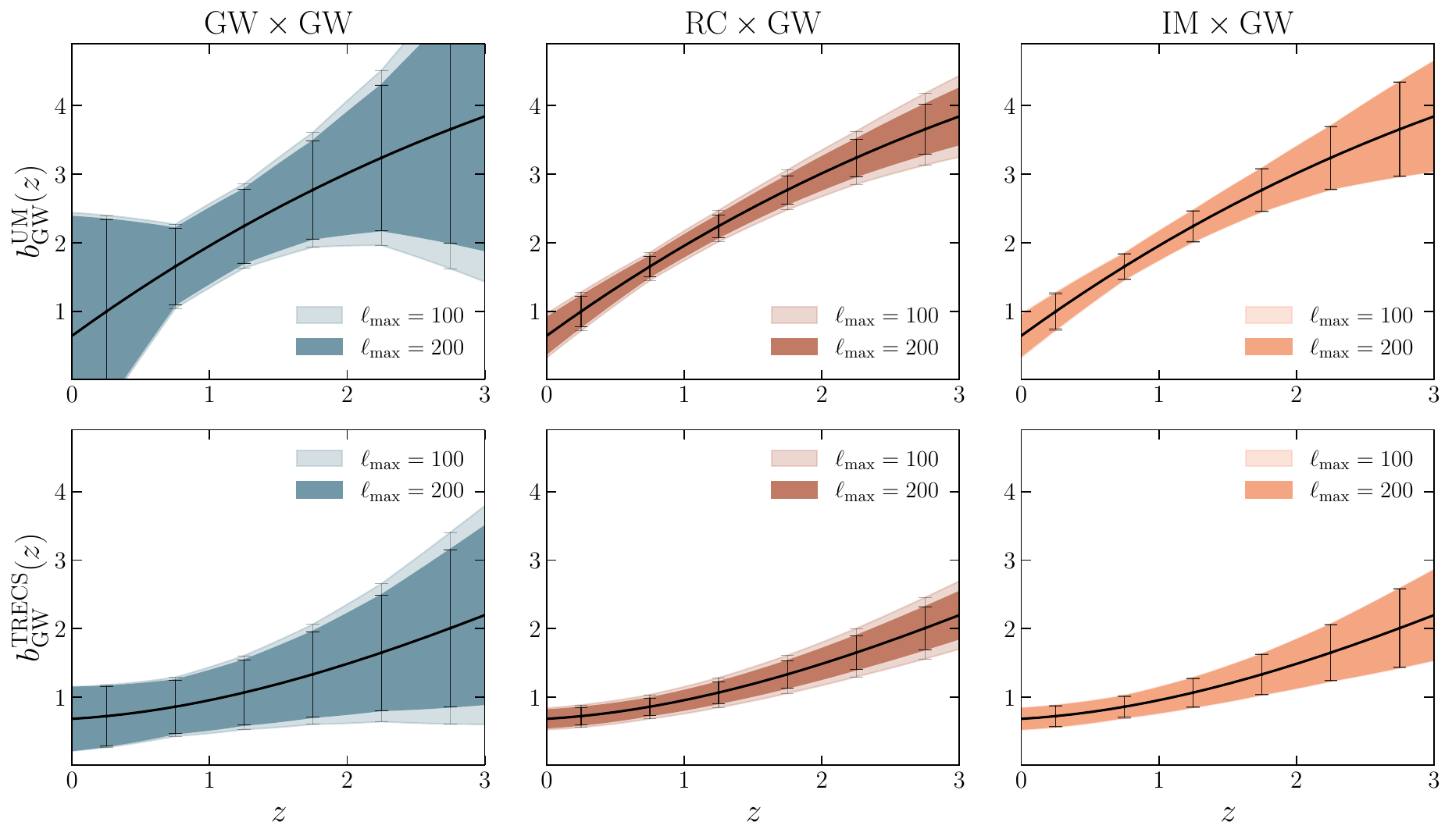}}
    \caption{Forecast constraints on the GW bias modeled using the UM (\textit{top panels}) and \texttt{T-RECS} (\textit{bottom panels}, analogous to Fig.~\ref{fig:sigma_bGW_ST_MT}) parametrizations, in Eq.~\eqref{eq:gw_um_bias} and~Eq.~\eqref{eq:gw_trecs_bias} respectively.
    }
\label{fig:compare_sigma_bGW_ST_MT}
\end{figure}

Clearly, changing the ABH bias model partially affects the capability of SKAO$\times$ET2CE to constrain its parameters. We re-run the Fisher forecast from Sec.~\ref{sec:gwbias_with_skao} using the UM parametrization and in Figure~\ref{fig:compare_sigma_bGW_ST_MT} we compare its results with what we previously obtained in Fig.~\ref{fig:sigma_bGW_ST_MT}. The bias expected when using the \texttt{T-RECS} prescription to link the SFR and host dark matter halo mass is smaller than the UM case, but the constraining power results comparable.

Furthermore, we verify that the outcomes in Sec.~\ref{sec:gwbias_PBH} are stable when using the UM bias parameterization instead of the \texttt{T-RECS} one, hence all these results are independent from the systematic uncertainties in the modeling of the SFR-$M_h$ relation.

\section*{Acknowledgments}

MB acknowledges that this article was produced while attending the PhD program in PhD in Space Science and Technology at the University of Trento, Cycle XXXIX, with the support of a scholarship financed by the Ministerial Decree no. 118 of 2nd March 2023, based on the NRRP - funded by the European Union - NextGenerationEU - Mission 4 "Education and Research", Component 1 "Enhancement of the offer of educational services: from nurseries to universities” - Investment 4.1 “Extension of the number of research doctorates and innovative doctorates for public administration and cultural heritage” - CUP E66E23000110001 and support by the Italian grant Project SPACE-IT-UP by the Italian Space Agency and Ministry of University and Research, Contract Number 2024-5-E.0.
SL thanks the Azrieli Foundation for support through an Azrieli International Postdoctoral Fellowship.
NB acknowledges support from the European Union's Horizon Europe research and innovation program under the Marie Sk\l{}odowska-Curie grant agreement no. 101207487 (GWSKY - Mapping the Universe with Gravitational Waves) and by PRD/ARPE 2022 ``Cosmology with Gravitational waves and Large Scale Structure - CosmoGraLSS''.
FS is supported by ICSC - Centro Nazionale di Ricerca in High Performance Computing, Big Data and Quantum Computing, funded by European Union - NextGenerationEU.
AR acknowledges funding from the Italian Ministry of University and Research (MIUR) through the ``Dipartimenti di eccellenza'' project ``Science of the Universe''.
ML acknowledges support by the MIUR Progetti di Ricerca di Rilevante Interesse Nazionale (PRIN) Bando 2022 - grant 20228RMX4A, funded by the European Union - Next generation EU, Mission 4, Component 1, CUP C53D23000940006.

\clearpage

\bibliographystyle{abbrvnat-maxbibnames4.bst}
\bibliography{bibliography}

\end{document}